\newcommand{\vecA}{\mbox{\boldmath$A$}}
\newcommand{\vece}{\mbox{{\boldmath$\hat{r}$}}}
\newcommand{\vecp}{\mbox{\boldmath$p$}}

\newcommand{\vecr}{\mbox{\boldmath$r$}}
\newcommand{\vecv}{\mbox{\boldmath$v$}}

\newcommand{\vecB}{\mbox{\boldmath$B$}}
\newcommand{\vecOmega}{\mbox{\boldmath$\Omega$}}
\documentclass[prb,showpacs]{revtex4}

\usepackage{graphicx}
\usepackage{dcolumn}
\usepackage{bm}

\begin{document}


\title{The Darwin-Breit magnetic interaction and superconductivity}

\author{Hanno Ess{\'e}n}
\affiliation{Department of Mechanics, Royal Institute of Technology\\ Stockholm SE-100 44, Sweden
}%
\author{Miguel C. N. Fiolhais}%
\affiliation{LIP-Coimbra, Department of Physics\\ University of Coimbra, Coimbra 3004-516, Portugal
}%

\date{2013 July 3}

\begin{abstract}
A number of facts indicating the relevance of the Darwin magnetic interaction energy in the superconducting phase are pointed out. The magnetic interaction term derived by Darwin is the same as the, so called, Breit term in relativistic quantum mechanics. While this term always is a small perturbation in few body systems it can be shown to be potentially dominating in systems of large numbers of electrons. It is therefore a natural candidate in the explanation of emergent phenomena---phenomena that only occur in sufficiently large systems. The dimensionless parameter that indicates the importance of the magnetic energy is the number of electrons times the classical electron radius divided by the size of the system. The number of electrons involved are only the electrons at the Fermi surface; electrons with lower energy cannot contribute to current density and thus not to the magnetic field.

The conventional understanding of superconductivity has always been problematic and no really reductionistic derivation exists. The idea that the inductive inertia, due to magnetism, is important in the explanation of superconductivity was first advanced by Frenkel and later brought up by Welker before it was prematurely discarded. So were theories involving Wigner crystallization. We speculate that the answer requires the combination of a Wigner lattice and the Darwin interaction. We point out that the Darwin interaction can be shown to have the right order of magnitude to explain the energy scales involved in normal superconductors. The London magnetic moment of rotating superconductors and the Meissner effect and their connection are discussed next. The London moment is shown to indicate that the number of electrons involved in the superconducting condensate is such that the Darwin interaction cannot be neglected.
\end{abstract}

\maketitle

\section{Introduction}
The standard understanding of superconductivity, at least of the original low temperature type, is, since the Bardeen-Cooper-Schrieffer (BCS) publication in 1957 \cite{SCbcs,SCgabovich&kuznetsov}, that it is due to phonons. These are assumed to produce an attractive interaction in a way analogous to the photon mediated electromagnetic interaction in vacuum. While this theory is microscopic in the sense that results are derived from the interaction between electrons it can not be considered as a fundamental theory since electrons do only have weak and electromagnetic interactions. There is no fundamental coupling constant for electron-phonon interactions, nor is the analogy between phonons and photons convincing for several reasons. The typical speed of phonons is the speed of sound and this speed is much lower that the typical speed of conduction electrons which is the Fermi velocity. Phonons are by definition acoustic and represent changes in mass density or velocity. Hence there is no electromagnetic effect associated with them, except possibly as some higher order effect. In fact most evidence on phonon coupling to the superconducting electrons can be interpreted as the destruction of the condensate by lattice oscillations. Consequently many theoretical physicists of accomplishment and integrity are still trying to understand the superconducting phase-transition from a more fundamental point of view (Hirsch \cite{SChirsch5,SChirsch1}, Vasiliev \cite{SCvasiliev}). We will also do this here. In particular we will point out that the magnetic interaction energy cannot be neglected.

We first point out how the magnetic interaction between electrons is very accurately described by the Darwin Lagrangian. To get some information on the quantum mechanics and statistical mechanics of the electrons one must proceed to the corresponding Hamiltonian--a non-trivial step. The various approximations to the Darwin Hamiltonian are then reviewed. After that we present some results that arise directly from electrodynamics for perfect conductors: the Meissner effect \cite{fiolhais&al,essen&fiolhais,fiolhais&essen} and the London moment \cite{essen05}. Finally there is a short account of how the magnetic interaction might work on electrons at the Fermi surface of metals.

\section{The Darwin approximation to electromagnetism}
In modern theoretical physics the Lagrangian formalism is considered the most basic. In 1920 Charles Galton Darwin published \cite{darwin} an approximation to the electromagnetic Lagrangian that removed the field degrees of freedom. It resulted from an expansion of the retarded potentials in $v/c$ to second order and contains only the positions and velocities of the charged particles of the system. In this approximation all electric and magnetic effects, except radiation, are well described. This Lagrangian is,
\begin{equation}\label{eq.L.ced}
L=T + \sum_a \frac{e_a}{2}
\left[\frac{\vecv_a}{c} \cdot \vecA(\vecr_a)- \phi(\vecr_a)
\right],
\end{equation}
where $T$ is the kinetic energy and $\phi$ is the Coulomb potential,
\begin{equation}
 \phi (\vecr_a) = \sum_{b(\neq a)}
\frac{e_b}{|\vecr_a -\vecr_b|} =\sum_{b(\neq a)}
\frac{e_b}{r_{ba}},
\end{equation}
and (hats are used for unit vectors, gaussian units are employed),
\begin{equation}
\label{eq.vec.pot.darw}
 \vecA (\vecr_a) = \sum_{b(\neq
a)} \frac{e_b [\vecv_b + (\vecv_b\cdot\vece_{ba})
\vece_{ba}] }{ 2c\, r_{ba}} ,
\end{equation}
is the form of the vector potential derived by Darwin. The form guarantees that the potential is divergence free.

To get the corresponding Hamiltonian one must perform the usual Legendre transform. This means that one must solve the equations
\begin{equation}
\label{eq.gen.moment.darw}
\vecp_a  = \frac{\partial L}{\partial \vecv_a},
\end{equation}
defining the generalized momenta in terms of the velocities, to get the velocities in terms of the momenta, and then insert the result in,
\begin{equation}
\label{eq.legendre.transf}
H=\sum_a \vecv_a \cdot \frac{\partial L}{\partial \vecv_a} -L.
\end{equation}
Even if only the non-relativistic kinetic energy, $T=\sum_a m_a \vecv_a^2/2,$ is used the solution of the equations (\ref{eq.gen.moment.darw}) involve the inversion of a large matrix. However, again using expansion in the parameter $v/c$ one obtains the Darwin Hamiltonian,
\begin{equation}
\label{eq.orig.darw.ham}
H = \sum_{a=1}^N \left[
\frac{\vecp_a^2}{2 m_a}+ \frac{e_a}{2}\phi_a(\vecr_a) -
\frac{e_a}{2m_a c} \vecp_a \cdot \vecA^p_a (\vecr_a)  \right] ,
\end{equation}
where,
\begin{equation}
\label{eq.darwin.A.ito.mom} \vecA^p_a (\vecr) = \sum_{b(\neq a)}^N
\frac{e_b [\vecp_b + (\vecp_b\cdot\vece_b) \vece_b] }{2m_b c
|\vecr-\vecr_b|} .
\end{equation}
This, original, Darwin Hamiltonian is correct if small $v/c$ really makes the Legendre inversion accurate. This is, however not necessarily the case for systems with large numbers of particles, and consequently this Hamiltonian may not be correct for superconductors and plasmas, or other macroscopic systems. It is possible to derive a second order correction, considered as a correction to the approximation in the Legendre transform \cite{essen96,essen97}.

By going to a continuum point of view and assuming a constant density of charged particles one can discover that the vector potential actually must be exponentially damped, like a Yukawa potential. This was discovered by Bethe and Fr\"ohlich \cite{bethe} in 1933, independently of Darwin's work. Later this has been studied more carefully \cite{essen99,essen&nordmark} by starting from the Darwin formalism and also by demanding that the vector potential remains divergence free, even when there is exponential damping. The length scale of the damping,
\begin{equation}
\label{eq.damp.length} \lambda = \frac{1}{\sqrt{4\pi r_{e} n}},
\end{equation}
is given by the same formula as that for the penetration depth in superconductors. Here $r_e =e^2 /mc^2 $ is the classical electron radius and $n$ the number density of electrons.

Bethe and Fr\"olich \cite{bethe} noted that if $n$ is the density of all electrons in some material one gets the surprising conclusion that magnetic field cannot penetrate into any material. I one instead interprets $n$ as the number density of the conduction electrons one still gets the strange conclusion that all metals would be perfectly diamagnetic. In superconductors $n$ is the density of electrons in the superconducting condensate and this is a much smaller number. So what determines the number density to be used in (\ref{eq.damp.length})? A clue comes from the fact that magnetism is intrinsically a retardation effect, and depends on the finite speed of light and the position and velocity of the charge. Thus it seems as if the only electrons that should be counted in $n$ are the electrons that have simultaneously defined position and velocity. This, of course, requires that they behave like semiclassical wave packets, instead of being in pure delocalized momentum states.

Consider Eqs.\ (\ref{eq.L.ced}) or (\ref{eq.orig.darw.ham}). We wish to estimate when the magnetic contributions to the Lagrangian or Hamiltonian no longer is a perturbation. If one simply assumes that the velocities or momenta are parallel it is easy to see that these terms become comparable to the kinetic energy term when
\begin{equation}
\label{eq.darw.num}
\nu=\frac{N r_e}{R}
\end{equation}
where $N$ is the number of electrons with correlated velocities (momenta), $r_e$ is the classical electron radius and $R$ is the typical distance between electrons (size of the system). When this dimensionless number $\nu$ no longer is small the magnetic term cannot be neglected. In the next section we'll see that for superconductors this number is 3.

\section{Electrodynamics and the London moment}
London  \cite{BKlondon} showed (see also \cite{SCbrady,SCcabrera,SCliu}), using his phenomenological theory
of superconductivity, that a superconducting sphere that rotates
with angular velocity
$\vecOmega$ will have an induced magnetic field (Gaussian units)
\begin{equation}
\label{eq.london.mag.field}
\vecB = \frac{2 mc}{|e|} \vecOmega
\end{equation}
in its interior. Here $m$ and $e$ are the mass and charge of
the electron. This prediction has been experimentally verified
with considerable accuracy and is equally true for high
temperature and heavy fermion superconductors
\cite{SCtate,SCsanzari}. With minor modifications it is also valid for
other axially symmetric shapes of the body, for example cylinders
or rings.

The most direct
way of understanding formula (\ref{eq.london.mag.field}) is,
in fact, quite simple. The superconducting electrons, which are
always found just inside the surface \cite{BKlondon}, are not
dragged by the positive ion lattice so when it starts to rotate
the superconducting electrons ignore this and remain in whatever
motion they prefer. This, however, means that there will be an
uncompensated motion of positive charge density on the surface
of the body. This surface charge density, $\sigma$, will, of
course, be the same  as the density of superconducting
electrons, but of opposite sign, and will produce the magnetic
field. Using this we can calculate the number, $N$, of
superconducting electrons.

It is well known that a rotating uniform surface charge density
will produce a uniform interior magnetic field in a sphere. If
this rotating surface charge density is $\sigma$, then the total
charge $Q$ is given by
\begin{equation}
Q=N |e|= 4\pi R^2 \sigma,
\end{equation}
and the resulting magnetic field in the interior is
\begin{equation}
\label{eq.int.mag.field.surface.dens}
\vecB = \frac{2}{3}\frac{Q}{cR} \vecOmega =
\frac{8\pi}{3}\frac{\sigma R}{c} \vecOmega ,
\end{equation}
where $R$ is the radius of the sphere (relevant formulas for the
calculation can be found in Ess\'en \cite{essen89}). Putting
$Q=N|e|$ and comparing this equation with
(\ref{eq.london.mag.field}) one finds that the number $N$ must be
given by $N=3 R m c^2/e^2 =3 R/r_e$. We thus find that the number defined in Eq.\ (\ref{eq.darw.num}) has the value
\begin{equation}
\label{eq.N.re.R.sup.cond.sphere}
\nu=\frac{N r_e}{R} =3 ,
\end{equation}
where $r_{e}$ is the classical electron radius, and $N$ the
number of electrons contributing to the supercurrent.
Note that here $R$ is the radius of the superconducting sphere, while in (\ref{eq.darw.num}) $R$ is the harmonic mean of all inter-particle distances, but the order of magnitude of $\nu$ is significant.

Note that the fundamental results above are universally true for all superconductors. How can this be if superconductivity is caused by some effective interaction with the lattice? A study of the connection between the London moment and the exclusion of a magnetic field from the interior in terms of the Darwin formalism can be found in \cite{essen05}. The reason that the current is only on the surface is that this minimizes the magnetic energy as shown in Fiolhais et al.\ \cite{fiolhais&al}.

\section{Magnetic energy of electrons on the Fermi surface}
Speculations of a magnetic origin of superconductivity originate quite early. Frenkel \cite{frenkel} advanced the
idea that inductive inertia causes the phenomenon. It was this idea that Bethe and Fr\"ohlich \cite{bethe} tried to disprove when they, to their consternation, discovered that all metals are perfectly diamagnetic, not just superconductors. Later the German physicist Welker \cite{SCwelker,SCwelker2} advanced similar ideas. Since Welker was not aware of the exponential damping effect he discovered that the magnetic interaction term in fact tends to diverge in large system.

Ess\'en \cite{essen95} studied the energy lowering effect of the original Darwin Hamiltonian (\ref{eq.orig.darw.ham}) if one assumes that only electrons on the Fermi surface, {\it i.e.} electrons that can behave semiclassically since many quantum states are available to them, contribute. One idea is that at low density of states (low density of semiclassical electrons) a Wigner \cite{SCwigner} lattice forms and that the electrons in this lattice then  move with correlated momenta.

Since the Fermi surface is two-dimensional one does not get the divergence that a three-dimensional distribution would give. Assuming that the electrons on the Fermi surface are distributed anisotropically in an optimal way one can the show that the Darwin approximation reduces the energy per electron at the Fermi surface by an amount,
\begin{equation}
\label{eq.lowedring.fermi.energy}
\Delta_D \approx 1.43 \cdot r_e\, k_F\, {\cal E}_F ,
\end{equation}
where $k_F$ is the Fermi wave number and ${\cal E}_F$ is the Fermi energy \cite{essen95}. In a typical metal this gives an energy lowering $\Delta_D \approx 10^{-4} {\cal E}_F$ which is in reasonable agreement with the smaller experimental energy gaps. This shows, if nothing else, that the Darwin magnetic energy ought to be a relevant quantity in superconductors.

\section{Conclusion}
We have reminded the reader of the fact that electrons interact also via magnetic fields. Macroscopically this corresponds to such phenomena as the attraction of parallel currents and the pinch effect in plasma physics, microscopically to the Breit \cite{breit29} interaction. The Darwin approximation and formalism may seem unfamiliar and exotic but it is in fact completely equivalent to ordinary electromagnetism when radiation can be neglected. Since several fundamental and universal facts of superconductivity, such as the London moment and the Meissner effect, are easily understood in terms of electrodynamics and magnetic energy, one is tempted to conjecture that this interaction energy also should be important in the microscopic realm. The order of magnitude estimate in the last section reinforces this impression.


----------

\end{document}